\documentclass[11pt,twoside]{article}
\usepackage{asp2004}
\usepackage{psfig}
\usepackage{epsf}
\usepackage{graphics}
\usepackage{lscape}
\def\edcomment#1{\iffalse\marginpar{\raggedright\sl#1\/}\else\relax\fi}
\reversemarginpar
\begin{document}
\title{Kinematics of Diffuse Ionized Gas Halos}
\author{R. J. Rand}
\affil{Dept. of Physics and Astronomy, University of New Mexico,
800 Yale Blvd, NE, Albuquerque, NM 87131}

\begin{abstract}

Existing long-slit spectral data for edge-on spiral galaxies suggesting that
their Diffuse Ionized Gas (DIG) halos rotate slower than their underlying
disks are summarized.  An attempt to characterize lagging halos using a model
of purely ballistic disk-halo flow is discussed, with the result that the
model fails badly for the lagging halo of NGC 891, but is somewhat more
successful for NGC 5775.  New two-dimensional kinematic data on the DIG halo
of NGC 4302 are presented, along with a preliminary analysis of its rotation.
Two-dimensional data on NGC 5775 and a preliminary analysis of its halo
rotation is discussed by Heald et al. (this volume).  The halo of NGC 4302
shows clear signs of lagging on its approaching side, but also strong
indications of peculiar kinematics.  The kinematics of the receding side 
are more complex.

\end{abstract}
\thispagestyle{plain}

\section{Introduction}

There is now much evidence for vertically extended layers of gas in spiral
galaxies that are most likely produced by a star-formation-driven flow from
the disk.  Such layers are seen in diffuse ionized gas (DIG;
e.g. \citealp*{rkh90}; \citealp*{d90}; \citealp*{hwr99}; \citealp*{rd03};
\citealp*{mv03}), 21-cm emission (e.g. \citealp*{ssh97}; \citealp{mw03};
Boomsma, this volume), and X-ray emission (e.g. \citealp{bh97};
\citealp{w01}), as well as radio continuum emission \citep*[e.g.][]{ddh94} and
dust absorption (e.g. \citealp*{hs99}; \citealp*{a00}; \citealp*{hs00}), among
other tracers.  Studies of the DIG layers strongly suggest that their
brightness and vertical extent depend on the level of underlying star
formation in the disk \citep[e.g.][]{r96,hwr99,rd03}.  This suggests that
while some gas may be infalling onto galaxy disks for the first time, the bulk
of what we have observed so far in these tracers is more likely the result of
a disk-halo flow.

While significant amounts of gas may typically participate in such flows,
little is known about whether the flows cause significant radial
redistribution of gas in the disk.  This will depend on the shape of the
galactic potential and the pressures and drag forces experienced by the halo
gas.  One way of attacking this question is to understand how gaseous halos
rotate relative to disks.  Information has begun to emerge on this question
through observations of atomic \citep[Boomsma, Fraternali, Osterloo, Matthews,
this volume]{ssh97,mw03} and ionized \citep{r97,r00,t01,mv03} gas, and this
has led to the first models of disk-halo cycling in which the flow is treated
in purely hydrostatic \citep[Ciotti, this volume]{b00} and ballistic
\citep{cbr02} limits.  In this paper, we present the observational data on DIG
halo rotation and summarize one attempt to explain the kinematics in term of a
purely ballistic flow.  An accompanying paper in this volume by Heald et
al. describes two-dimensional kinematic data on NGC 5775.

\section{Long-slit Spectroscopy of NGC 891 and NGC 5775}

Evidence for lagging DIG halos in NGC 891 and NGC 5775 has been presented by
\citet{r97,r00} and \citet{t01}.  A long-slit spectrum oriented perpendicular
to the plane of NGC 891, centered at 100'' from the nucleus (5 kpc at the
assumed distance of 9.5 Mpc) on the approaching side of the galaxy, shows
increasing heliocentric mean velocities (Figure 1), consistent with a
declining rotation curve (the initial decrease within 1 kpc of the midplane is
an effect of dust absorption).  The change in observed velocity is some 20--30
km s$^{-1}$ from 1 to 4 kpc off the plane.  A lag is also seen in HI, but the
falloff is significantly steeper at this location \citep[Fraternali, this
volume]{ssh97}.

\begin{figure}[!ht]
\plotfiddle{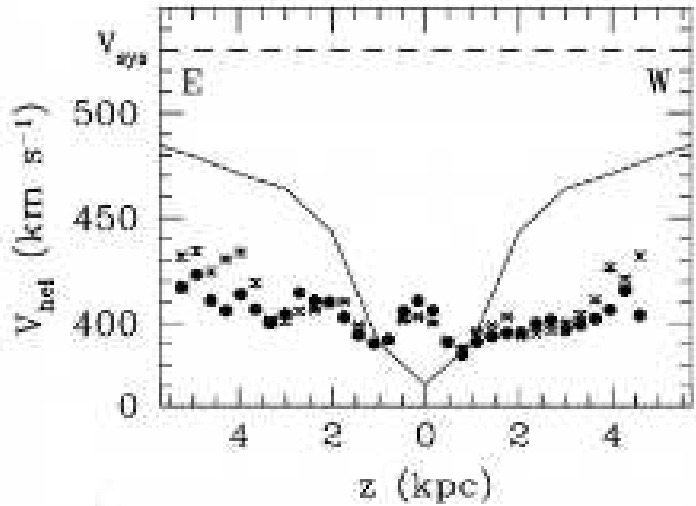}{3truein}{0}{120}{120}{-380}{-375}
\caption{H$\alpha$ (crosses) and [N$\,$ II]$\lambda$6583 (filled circles)
heliocentric line centroids are shown for a long-slit spectrum of NGC 891
centered at 100'' northeast of the nucleus and oriented perpendicular to the
plane.  The systemic velocity of $V_{sys} = 530$ km s$^{-1}$ is indicated by
the dashed line.  The solid curve is the prediction of a ballistic model of
disk-halo cycling.}
\end{figure}

\begin{figure}[!ht]
\plotfiddle{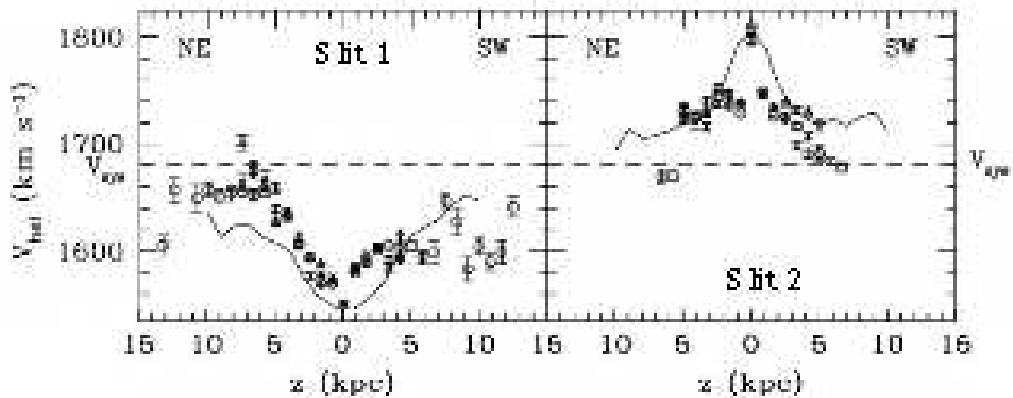}{2.5truein}{0}{120}{120}{-370}{-380}
\caption{[N$\,$II]$\lambda$6583 (open circles), [S$\,$II]$\lambda$6716
(triangles), and [S$\,$II]$\lambda$ 6731 (crosses) heliocentric line centroids
are shown for long-slit spectra of NGC 5775 centered at 32'' northwest (Slit
1) and 20'' southeast (Slit 2) of the nucleus and oriented perpendicular to
the plane.  The systemic velocity of $V_{sys} = 1680$ km s$^{-1}$ is indicated
by the dashed line.  The solid curve is the prediction of a ballistic model of
disk-halo cycling.}
\end{figure}

A more dramatic example is NGC 5775.  This is an interacting galaxy
\citep[e.g.][]{i94} with a very bright and extended DIG halo, although more
concentrated into large shell, filamentary and patchy structures compared to
the halo of NGC 891 \citep[also in HI; see][]{l01}.  At some level, the halo
kinematics may be modified by the interaction.  Mean heliocentric velocities
of DIG for two long-slits (Slit 1 on the approaching side and Slit 2 on the
receding side) are shown in Figure 2.  Forbidden-line velocities are shown
rather than H$\alpha$ velocities because faint emission from the latter is
confused by imperfect sky-line subtraction.  Velocities are measurable up to
about 12 kpc from the plane in Slit 1.  The trend for both slits is consistent
with a lagging halo, with velocities approaching or reaching $V_{sys}$ at the
largest heights.  Further long-slit data by \citet{t01}, near the location of
Slit 2, show a similar trend.

\section{A Ballistic Model of Disk-Halo Flow}

Assuming that the extraplanar gas in these galaxies is a result of disk-halo
circulation, what can the lags tell us about the nature of the flow?  The
discoveries are new enough so that models of the circulation are not yet very
sophisticated.  In one limit, one can consider a purely hydrostatic continuum
fluid disk \citep[ Ciotti, this volume]{b00}.  In the other extreme, one can
assume a purely ballistic flow with no pressure, tension, drag, or cloud
interactions \citep{cbr02}.  Each approach should shed light on the true
nature of the flow.

In the ballistic model of \citet{cbr02}, clouds are launched from the disk
with some vertical velocity chosen from a uniform, random distribution of
values between zero and some maximum velocity $V_{kick}$.  The initial
location of the clouds are chosen from an exponential probability distribution
in $R$ and a narrow Gaussian distribution in $z$.  The potential used for most
simulations is from \citet{w95}.  As the clouds rise they migrate outward in
radius due to the weaker radial component of the potential they experience;
they rotate slower as a consequence of conservation of angular momentum.  The
most important parameter governing the extent and kinematics of the halo
clouds is the ratio of $V_{kick}$ to the circular rotation speed $V_{circ}$.
This ratio is chosen to match the vertical scale-height of the emission for
the halo in question.  Figure 3, showing meriodonal plots ($z$ vs. $R$) for
clouds launched at radii of 4, 8, 12 and 16 kpc for $V_{kick}$/$V_{circ}=0.5$,
demonstrates the outward migration and reduction in azimuthal velocities of
the halo clouds.

\begin{figure}[!ht]
\plotone{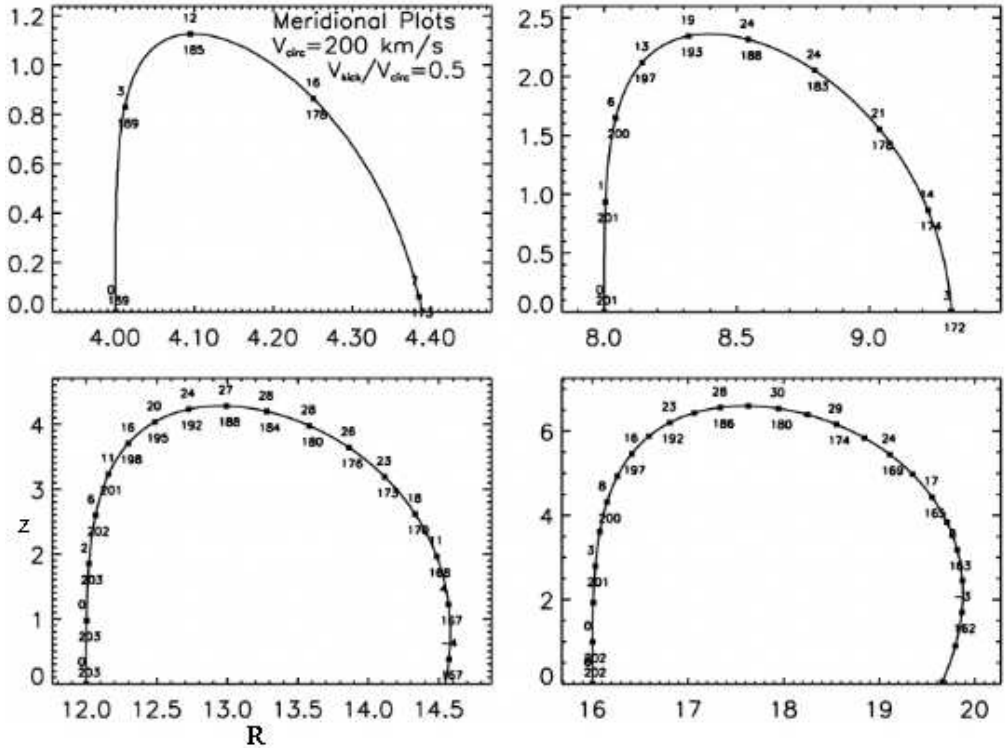}
\caption{Meridional plots from the model of \citet{cbr02} showing
sample orbits for a case with $V_{circ}$ = 200 km s$^{-1}$ and $V_{kick}$ =
100 km s$^{-1}$. In each case, the horizontal axis is radius in kpc,
and the vertical axis is height in kpc.  Points note the
position of the particle at 20 Myr intervals. The number above each
point indicates the outward radial velocity; the number below gives the
azimuthal velocity.}
\end{figure}

Figures 1 and 2 show how well the predicted mean heliocentric velocities match
the data for NGC 891 and NGC 5775.  The model badly overpredicts the lag in
NGC 891 but fares somewhat better for NGC 5775, although the observed
velocities near systemic cannot be reproduced.  In NGC 891, the outward
migration and resultant lag is presumably mitigated by some
(magneto)hydrodynamic effect, such as a halo gas pressure increasing radially
outward or a magnetic or viscous coupling to the disk \citep[Ciotti, this
volume]{b00}.  In NGC 5775, with its apparently more active disk-halo flow
(the emission is brighter, the gas scale-height larger, and the morphology
more filamentary than in NGC 891), the ballistic approximation may work
somewhat better.

\section{New 2-D Kinematic Data for NGC 4302}

NGC 4302 is paired with another spiral NGC 4298, although there are no known
signs of interaction.  The galaxy appears extremely edge-on.  The extraplanar
DIG is concentrated above the inner disk, as for NGC 891, but several times
fainter at comparable heights \citep{r97}.  We have observed NGC 4302 with two
pointings of the Sparsepak fiber-optic array \citep{b04} on the WIYN
telescope.  Figure 4 shows the spatial sampling of the fibers.  Total
integration times are 380 and 410 minutes for the northern and southern
pointings, respectively.  The velocity resolution is 30 km s$^{-1}$, and the
bright lines covered include H$\alpha$, [N$\,$II]$\lambda\lambda 6548, 6583$
and [S$\,$II]$\lambda\lambda 6716, 6731$.  The results presented here are
preliminary, and based only on mean velocities, which are indicative of, but
do not accurately represent, rotation speeds.

\begin{figure}[!ht]
\plotfiddle{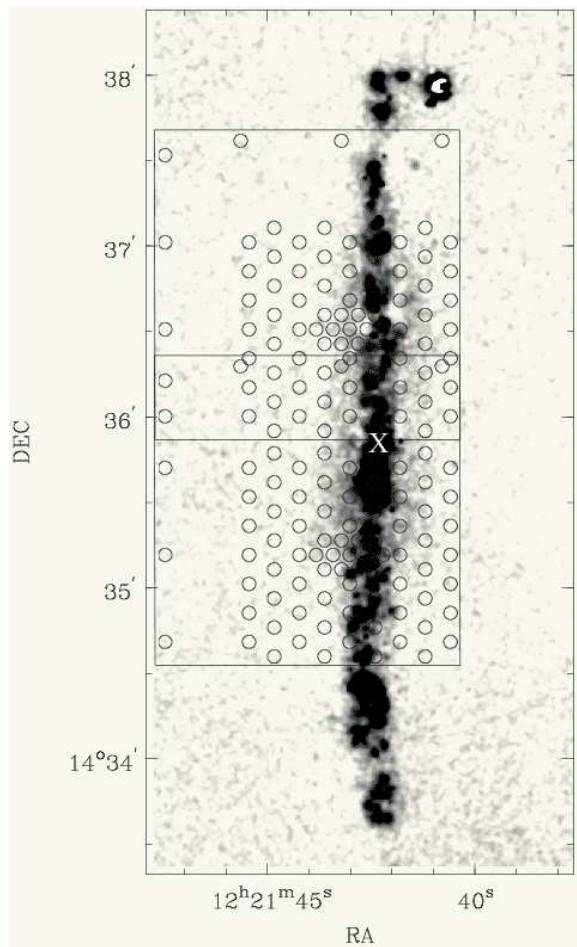}{5truein}{0}{45}{45}{-120}{0}
\caption{H$\alpha$ image of NGC 4302 from \citep{r97} showing the two
pointings of the Sparsepak array.}
\end{figure}

\begin{figure}[!ht]
\plotfiddle{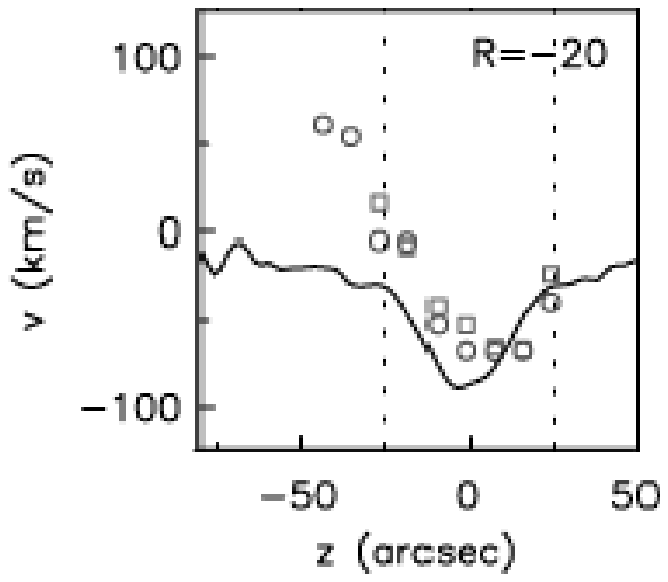}{4truein}{0}{100}{100}{-405}{-200}
\plotfiddle{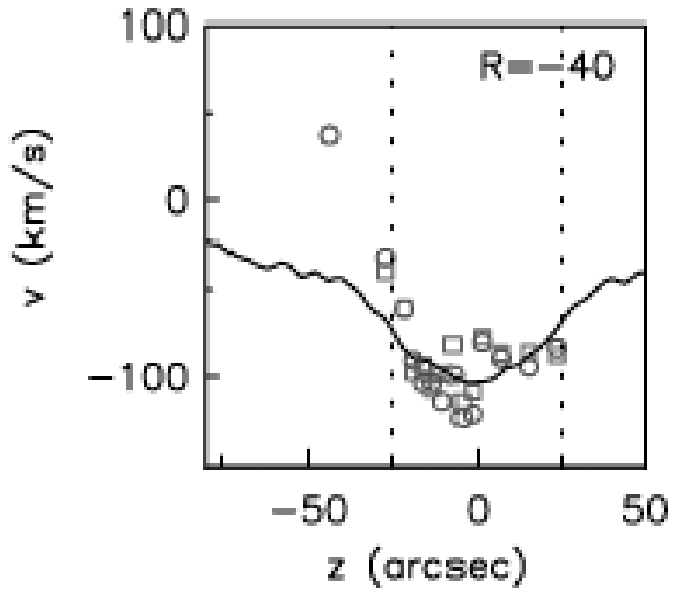}{0truein}{0}{100}{100}{-215}{-172}
\plotfiddle{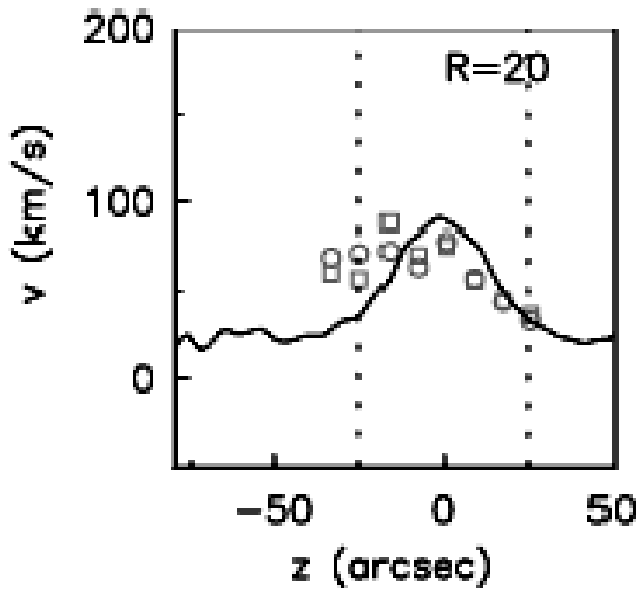}{0truein}{0}{100}{100}{-404}{-330}
\plotfiddle{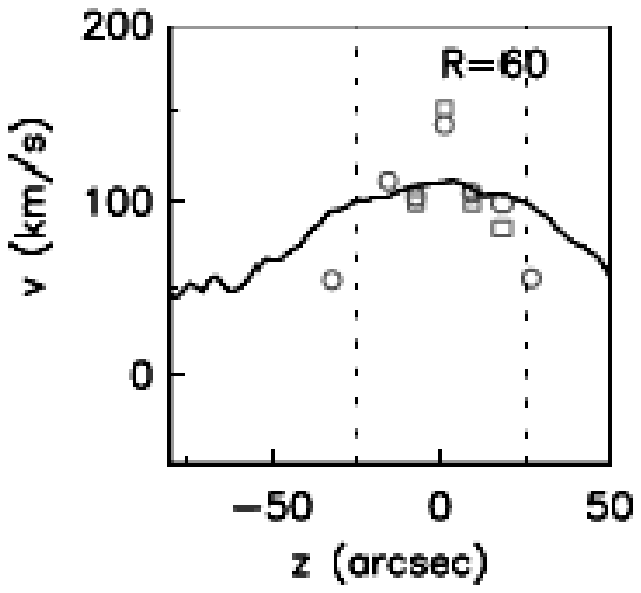}{0truein}{0}{100}{100}{-210}{-305}
\caption{Velocities of the H$\alpha$ (squares) and [N$\,$II]$\lambda 6583$
(circles) emission as a function of height from the plane for four distances
(in arcsec) from the minor axis of NGC 4302.}
\end{figure}

Figure 5 shows velocities of the H$\alpha$ and [N$\,$II]$\lambda 6583$ lines
as a function of height from the midplane for four representative distances,
R, from the minor axis (negative values of R correspond to the approaching
side and positive values to the receding side).  A preliminary run of the
ballistic model from \citet{cbr02} is also shown in each panel.  The ratio
$V_{kick}$/$V_{circ}$ is 0.8, but model parameter space has not yet been
explored.  In general, the approaching side of the galaxy suggests a
reasonably well-behaved lagging halo, which can be fairly matched by the
ballistic model, but at heights larger than about 25'' (2 kpc) mean velocities
become very close to systemic and in many cases become greater than systemic,
suggesting a disturbance or possibly counterrotation.  The kinematics of the
receding side are more complex, with much more scatter in the velocities (the
R=20'' panel being typical) and a lagging halo signature apparent in only a
few locations, such as at R=60" as shown here.

\section{Conclusions and Future Work}

There is clear evidence that ionized gas halos rotate slower than the
underlying disks.  The falloffs in rotation speed are not generally matched by
a ballistic model of disk-halo flow, indicating that pressures and/or drag
forces must affect rotation significantly.  The overpredicted lag in NGC 891
suggests either a halo gas pressure that increases with radius or a magnetic
or viscous drag with the faster disk.  The model is somewhat more successful
in NGC 5775 and NGC 4302, but cannot reproduce velocities near the systemic
value at heights of several kpc in the former, while the latter shows several
peculiarities in its halo kinematics.  However, while mean velocities give an
indication of the lagging effect, they cannot be used to determine with
accuracy the decline of the rotation speed with height.  It is therefore
necessary to produce kinematic models that match the full spatial and spectral
information available in two-dimensional kinematic data.  We have begun this
process using observations of NGC 5775 in the H$\alpha$ line using the Taurus
Imaging Fabry-Perot Interferometer on the Anglo-Australian Telescope (Heald et
al., this volume).  This is a difficult process, especially given the 86$\deg$
inclination of this galaxy.  Dust further complicates the analysis.  We will
also apply this analysis to the data on NGC 4302 described above.

\acknowledgements{This material is based on work partially supported by the
National Science Foundation under Grant No. AST 99-86113.}

{}

\end{document}